\documentclass[aps,twocolumn,showpacs,nofootinbib,superscriptaddress]{revtex4-1}

\usepackage{color} 

\definecolor{green}{rgb}{0,.5,0}

\definecolor{red}{rgb}{1,0,0}
\newcommand{\red}[1]{{#1}}
\usepackage{graphicx}
\usepackage[subfigure]{graphfig}
\usepackage{epsfig}
\usepackage{dcolumn}
\usepackage{amsmath}
\usepackage{latexsym}
\usepackage{float}

\def\be{\begin{equation}}
\def\ee{\end{equation}}
\def\bea{\begin{eqnarray}}
\def\eea{\end{eqnarray}}

\begin{document}

\title{Detecting the flavor content of the vacuum using the Dirac operator spectrum}

\author{
\includegraphics[scale=0.15]{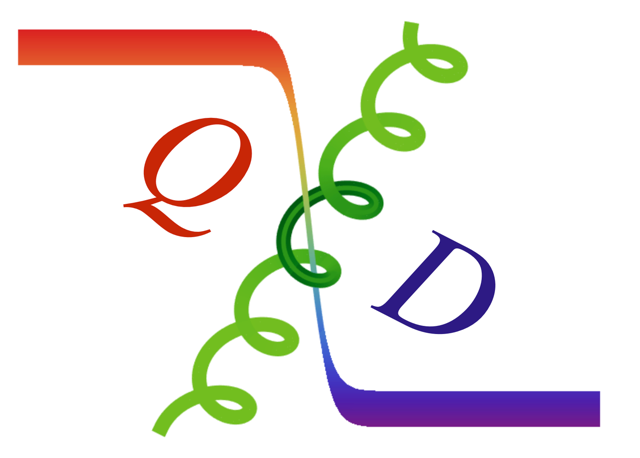}\\
Jian Liang}
\email{jianliang@scnu.edu.cn}
\affiliation{
Key Laboratory of Atomic and Subatomic Structure and Quantum Control (MOE), Guangdong Basic Research Center of Excellence for Structure and Fundamental Interactions of Matter, Institute of Quantum Matter, South China Normal University, Guangzhou 510006, China}
\affiliation{
Guangdong-Hong Kong Joint Laboratory of Quantum Matter, Guangdong Provincial Key Laboratory of Nuclear Science, Southern Nuclear Science Computing Center, South China Normal University, Guangzhou 510006, China}
\affiliation{Department of Physics and Astronomy, University of Kentucky, Lexington, KY 40506, USA}

\author{Andrei Alexandru}
\affiliation{Department of Physics, The George Washington University, Washington, DC 20052, USA}

\author{ Yu-Jiang Bi}
\affiliation{Institute of High Energy Physics, Chinese Academy of Sciences, Beijing 100049, China}

\author{Terrence Draper}
\affiliation{Department of Physics and Astronomy, University of Kentucky, Lexington, KY 40506, USA}

\author{Keh-Fei Liu}
\affiliation{Department of Physics and Astronomy, University of Kentucky, Lexington, KY 40506, USA}

\author{Yi-Bo Yang}
\email{ybyang@itp.ac.cn}
\affiliation{\mbox{CAS Key Laboratory of Theoretical Physics, Institute of Theoretical Physics, Chinese Academy of Sciences, Beijing 100190, China}}
\affiliation{\mbox{School of Fundamental Physics and Mathematical Sciences, Hangzhou Institute for Advanced Study, UCAS, Hangzhou 310024, China}}
\affiliation{\mbox{International Centre for Theoretical Physics Asia-Pacific, Beijing/Hangzhou, China}
}
\collaboration{$\chi$QCD Collaboration}

\begin{abstract}

We compute the overlap Dirac spectrum on three gauge ensembles generated using $2+1$-flavor domain wall fermions. 
The three ensembles have different lattice spacings and two of them have quark masses tuned to the physical point. 
The spectral density is determined up to $\lambda\sim$100 MeV with subpercentage statistical uncertainty.
We find that the density is close to a constant below $\lambda\sim$ 20 MeV 
as predicted by chiral perturbative theory ($\chi$PT), 
and then increases linearly due to the strange quark mass. 
By fitting to the next-to-leading order $\chi$PT form and using the non-perturbative RI/MOM renormalization, 
the $\rm SU(2)$ (keeping the strange quark mass at the physical point) 
and $\rm SU(3)$ chiral condensates at $\overline{\textrm{MS}}$ 2 GeV are determined to be
$\Sigma=(265.4(0.5)(4.2)\ \textrm{MeV})^3$ and $\Sigma_0=(234.3(0.5)(25.8)\ \textrm{MeV})^3$,
respectively.
The pion decay constants are also determined to be $F=84.1(1.9)(8.0)$ and $F_0=58.6(0.5)(10.0)$ MeV. 
The systematic errors are carefully estimated including
the effects of fitting ranges and
the uncertainty of low-energy constant $L_6$.
We also show that
one can resolve the sea flavor content of the sea quarks and constrain their masses with {$\sim10\%-20\%$} statistical uncertainties using the Dirac spectral density.

\end{abstract}

\maketitle

\section{Introduction}
The QCD vacuum includes both quarks and gluons. 
A gluon interacts with itself directly, but a quark can only interact with other quarks indirectly through gluons. 
So a fundamental issue is how a quark's properties depend upon the flavors of virtual quarks in the vacuum.
The quark Dirac operator $D\!\!\!\!/=(\partial_{\mu}-igA_{\mu})\gamma_{\mu}$ is an efficient tool to address this question, 
as the quark flavors in the QCD vacuum (the ``sea" quarks) just affect $D\!\!\!\!/$\ through the gluon field $A_{\mu}$. Based on the Banks-Casher relation~\cite{Banks:1979yr},
the near-zero spectrum density $\rho(\lambda\sim0)$ of $D\!\!\!\!/$\ \ is proportional to the vacuum chiral condensate $\langle \bar{\psi}\psi\rangle$, 
which is the order parameter of spontaneous chiral symmetry breaking. 
By treating $\lambda$ as an imaginary quark mass~\cite{Damgaard:2008zs}, chiral perturbative theory ($\chi$PT) reproduces the Banks-Casher relation as its leading order approximation, and provides a quantitative next-to-leading order (NLO) correction to $\rho(\lambda)$ due to the masses of different quark flavors in the vacuum, in the small quark mass region where $\chi$PT is valid. 
But such a $\chi$PT calculation becomes quite complicated beyond NLO
when the finite volume effects are taken into account, and its predictability is limited as there are many parameters in next-to-next-leading order (NNLO).

On the other hand, non-perturbative lattice QCD can provide first-principles information about $\rho(\lambda)$ which has been shown to be renormalizable~\cite{DelDebbio:2005qa,Giusti:2008vb}, while it is still challenging to be precise and accurate enough to provide the flavor information of the vacuum. 
First of all, one has to use the overlap fermions~\cite{Chiu:1998gp,Liu:2002qu} or the projected Domain-wall fermions~\cite{Brower:2012vk} to obtain $\rho(\lambda)$ accurately at $\lambda\sim0$~\cite{Fukaya:2010na,Fukaya:2009fh,Cossu:2016eqs},
but at the expense of $\cal O$(10) times or more computer time than that of the standard Wilson-like discretization of $D\!\!\!\!/$ that breaks chiral symmetry explicitly.
At the same time, one needs to solve the smallest eigenpairs (eigenvalues and the corresponding eigenvectors) of $D\!\!\!\!/$ to extract $\rho(\lambda)$ precisely. This is $\cal O$(100) times more expensive than the cost of a standard quark propagator calculation. 
Thus the stochastic method has been proposed to provide some estimates of $\rho(\lambda)$~\cite{Cichy:2013gja,Engel:2014cka,Engel:2014eea,Cossu:2016eqs}.

During investigations in the last decade, the low-lying $D\!\!\!\!/$ eigenpairs have been found to be extremely beneficial in lattice QCD calculations of correlators to improve the signal-to-noise ratio using the low-mode substitution (LMS) algorithm for the noise grid source~\cite{Li:2010pw,Yang:2015zja,Liang:2016fgy}. 
These low modes tend to saturate the long-range parts of hadron correlators and are also useful in calculating quark loops together with the noise estimation for the high modes~\cite{Gong:2013vja}.  
In this work, we solve the low-lying eigenpairs of the chiral fermion directly, and then present the result of the spectrum density $\rho(\lambda)$ with the smallest statistical uncertainty to date on $2+1$-flavor physical quark mass ensembles at three lattice spacings, and determine the chiral condensate and the pion decay constants for both the $\rm SU(2)$ and $\rm SU(3)$ cases.
We also try to extract other sea information such as the quark masses.

\section{Methodology and Numerical Setup}

The overlap fermion was proposed in Refs.~\cite{Chiu:1998eu,Liu:2002qu} to 
construct a discretized fermion action satisfying the Ginsparg-Wilson relation 
$D_{\textrm{ov}}\gamma_5+\gamma_5D_{\textrm{ov}}=\frac{a}{M_0}D_{\textrm{ov}}\gamma_5D_{\textrm{ov}}$~\cite{Ginsparg:1981bj}, 
\begin{equation}
D_{\textrm{ov}}=M_0\Big(1+\gamma_5\epsilon\big(H_{\rm w}(-M_0)\big)\Big),
\end{equation}
where $\epsilon(H_{\rm w})=\frac{H_{\rm w}}{\sqrt{H_{\rm w}^2}}$ is the matrix sign function, $H_{\rm w}(-M_0)=\gamma_5D_{\rm w}(-M_0)$ and $D_{\rm w}$ is the Wilson Dirac operator with a negative mass parameter such as $M_0=1.5$ to avoid the singularity in $1/D_{\rm w}$. $\epsilon(H)$ can be decomposed into a combination of the small and large eigenvalue regions,
\begin{align}
\epsilon(H_{\rm w})=&\sum_{\lambda_{{\rm H},i}<\lambda^{\rm cut}_{\rm H}}\frac{\lambda_{{\rm H},i}}{|\lambda_{{\rm H},i}|}\textbf{v}_{{\rm H},i}(\textbf{v}_{{\rm H},i})^{\dagger}\nonumber\\
&+(1-\sum_{\lambda_{{\rm H},i}<\lambda^{\rm cut}_{\rm H}}\textbf{v}_{{\rm H},i}(\textbf{v}_{{\rm H},i})^{\dagger}){\sum^N_{j}d_j H_{\rm w}^{2j+1}},
\end{align}
in which the eigenpair $(\lambda_{{\rm H},i},\ \textbf{v}_{{\rm H},i})$ satisfies the relation $H_{\rm w}\textbf{v}_{{\rm H},i}=\lambda_{{\rm H},i}\textbf{v}_{{\rm H},i}$, 
and $d_{j=0,1,2,\dots,N}$ are the Chebyshev polynomial coefficients to 
approximate $\epsilon(H_{\textrm w})$ 
for all the eigenvalues larger than a given cutoff $\lambda^{\rm cut}_{\rm H}$ 
with a given accuracy~\cite{Giusti:2002sm}. The computational cost of $D_{\rm ov}$ 
is proportional to the polynomial order $N$ and one can increase $\lambda^{\rm cut}_{\rm H}$ 
to reduce $N$ until the $\lambda_{{\rm H},i}$ become very dense for $\lambda^{\rm cut}_{\rm H}\sim0.2/a$, 
where $a$ is the lattice spacing. 
Even for the optimal $\lambda^{\rm cut}_{\rm H}$ we still have $N\sim 100$; 
thus, such a large $N$ makes the computational cost of $D_{\rm ov}$ two orders of magnitude higher than that of $D_{\rm w}$. 

The chiral Dirac operator can be defined through $D_{\rm ov}$, 
\begin{align}
D_{\rm c}=\frac{D_{\rm ov}}{1-\frac{1}{2M_0}D_{\rm ov}}=\frac{M_0}{2}\frac{1+\gamma_5\epsilon(\gamma_5D_{\rm w}(M_0))}{1-\gamma_5\epsilon(\gamma_5D_{\rm w}(M_0))},
\end{align}
and satisfies the same commutation relation $D_{\rm c}\gamma_5+\gamma_5D_{\rm c}=0$ as that of $D\!\!\!\!/\ $ in the continuum. Each eigenvector of $D_{\rm c}$ is exactly the same as that of $D_{\rm ov}$, and the eigenvalue $i\lambda_{\rm c}$ of  $D_{\rm c}$ can be obtained from that of $D_{\rm ov}$ by the relation $i\lambda_{\rm c}=\frac{\lambda_{\rm ov}}{1-\frac{1}{2M_0}\lambda_{\rm ov}}$ and is purely imaginary. Using the Arnoldi factorization algorithm~\cite{arnoldi1951}, one can obtain the low-lying eigenpairs of $D_{\rm ov}$, and thus those of $D_{\rm c}$. Both $D_{\rm ov}$ and $D_{\rm c}$ are similar to $D_{\rm w}(0)$ at the continuum limit while providing a proper ultraviolet cutoff to preserve chiral symmetry. {Then we can define the eigenvalue density of $D_{\rm c}$ as $\rho(\lambda_{\rm c})\equiv\frac{\partial n(\lambda_{\rm c})}{V \partial \lambda_{\rm c}}$, where $V=L^3\times T$ is the 4-D volume and $n(\lambda_{\rm c})$ is the number of eigenvalues in the range $(0,i\lambda_{\rm c}]$. Since $\rho(\lambda_c)$} is so far the best approximation of the spectrum density on a discretized lattice, we will use $\lambda_{\rm c}$ in the following discussion and omit the subscript. 

Based on the partially quenched chiral perturbative theory (PQ$\chi$PT) which can accommodate the valence quark masses being different from the sea quark masses, one can derive the formula to describe $\langle \bar{\psi}\psi\rangle$ of the valence quark as a function of both valence and sea quark masses. In PQ$\chi$PT~\cite{Damgaard:2008zs}, the density $\rho(\lambda)$ of the low-lying eigenvalues can be expressed as the chiral condensate $\langle \bar{\psi}\psi\rangle$ with virtual quark mass $i\lambda$ in a finite volume $V=L^3\times T$,
\begin{align}\label{eq:def}
\rho(\lambda,V)=&\lim_{\epsilon \rightarrow 0}\frac{1}{2\pi}\big(\langle \bar{\psi}\psi\rangle|_{m_v=i\lambda-\epsilon}-\langle \bar{\psi}\psi\rangle|_{m_v=i\lambda+\epsilon}\big)\nonumber\\
=&\frac{\Sigma}{\pi}\textrm{Re}\left[Z_v(i\lambda,m_{q}^{\rm sea}) \hat{\Sigma}^{\rm PQ}(i\lambda V,m_{q} V)\right],
\end{align}
where $\Sigma\equiv-\lim_{m_q\to0}\lim_{V\to\infty}\langle \bar{\psi}\psi\rangle$ is the chiral condensate in the chiral limit of $N_f$ flavors, and $m_q$ is the light or strange sea quark mass.
The standard NLO correction 
\begin{align}\label{eq:Z_v}
Z_v=&\ 1+\beta_1(L,T)\frac{N_f^2-1}{N_f}\frac{1}{F^2\sqrt{V}}+\frac{N_f^2-4}{N_f}\frac{\lambda \Sigma}{32\pi F^4}\nonumber\\
&+{\cal O}(m^{\rm sea}_{q},\lambda^2)
\end{align}
has leading order finite-volume correction which is $\sim$4\% with $N_f=2$ for a $(5.5\ {\rm fm})^4$ lattice~\cite{Hasenfratz:1989pk}, and the $\lambda$-dependent finite volume correction
\begin{align}
 &\hat{\Sigma}^{\rm PQ}(i\lambda V,m_{q} V) =1-\frac{1}{\Sigma V}(\sum_{q=u,d,s} \frac{1}{i\lambda+m_{q}}\nonumber\\
 &\quad\quad+\frac{1}{2\lambda^2\sum_{q=u,d,s}\frac{1}{m_{q}}}) +{\cal O}\big((m_{q})^2,\lambda^2,\frac{1}{V^2}\big),
\end{align}
is suppressed at large $\lambda$. It is important to note that $Z_v$ can differ from 1 by $\sim$30\% or more if we keep the strange quark mass at the physical point, and it makes the $\Sigma$ in the $N_f$ = 2 and 3 chiral limits quite different from each other. We skip the complete NLO expression of $\rho(\lambda,V)$ since it is very lengthy, especially in the $N_f$ = 2+1 case, and the interested reader can find it in Ref.~\cite{Damgaard:2008zs}. 

\begin{table}                   
\caption{Information of the RBC ensembles~\cite{Blum:2014tka,Mawhinney:2019cuc} used in this calculation. The pion and kaon masses are in unit of MeV.}  

\begin{tabular}{l l l c l c | c}                                                                
\text{Symbol} & $L^3 \times T$  &  $a$ (fm)   & $m_\pi$ & $m_K$ & $Z_S$ &  $N_\text{cfg}$\\
\hline   
48I &$48^3\times\ 96$& 0.1141(2) & 139 & 499 & {1.117(1)(17)} &303  \\
\hline    
64I &$64^3\times128$& 0.0837(2) & 139 & 508 & {1.028(1)(13)} & 304 \\                                                         
\hline
48IF    & $48^3 \times\ 96$ & 0.0711(3) & 234 &564 & {0.989(1)(17)} & 185\\
\hline
32IF    & $32^3 \times\ 64$ & 0.0626(4) & 371 &558 & {0.951(2)(12)} & 50\\
\end{tabular}  
\label{tab:ensemble}           
\end{table}   

We list the four ensembles we used in this work in Tab.~\ref{tab:ensemble}, including two ensembles at the physical light and strange quark masses. \red{Those ensembles were generated by the RBC/UKQCD collobaration using the 2+1 flavor Domain wall fermion and Iwasaki gauge action.}  
We will use just the first three ensembles to extract the spectrum density $\rho(\lambda)$, 
and the last ensemble with the finest lattice spacing will just be used to show the lattice spacing dependence of the scalar renormalization constant. 
With one step of HYP smearing~\cite{Hasenfratz:2001tw}, we need 800 $H_{\rm w}$ eigenpairs and 1000 pairs of $D_{\rm c}$ eigenpairs to reach the upper bounds 0.158 and $\pm0.056i$ respectively on the largest 64I ensemble, while the same upper bounds can be reached with only $\sim$180 $H_{\rm w}$ eigenpairs and $\sim$60 pairs of $D_{\rm c}$ eigenpairs on a small $24^3\times 64$ gauge ensemble at lattice spacing 0.1105(2) fm.

\section{Renormalization}\label{sec:mom-vs-smom}

\begin{figure}[htbp]
	\centering
	\includegraphics[width=0.48\textwidth]{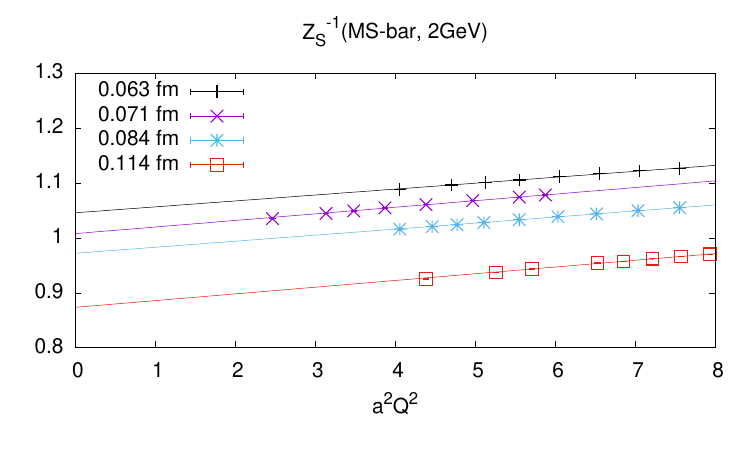}
	\caption{The inverse of the scalar current renormalization constant at $\overline{\textrm{MS}}$ 2 GeV and different lattice spacings, as a function of RI/MOM off-shell scale $Q^2$. The $a^2Q^2$ dependencies in the figure for different lattice spacings are similar, which means that the residual $Q^2$ dependence vanishes in the continuum limit.}
	\label{fig:Zm}
\end{figure}

The bare chiral condensate and $\rho(\lambda)$ we obtained are under the lattice regularization and require renormalization. To renormalize the scalar quark bi-linear operator, we use the regularization independent momentum subtraction (RI/MOM) scheme~\cite{Martinelli:1994ty,Bi:2017ybi} under the Landau gauge, and further convert the result into $\overline{\textrm{MS}}$ 2 GeV as $\langle \bar{\psi}\psi\rangle^r=Z_S\langle \bar{\psi}\psi\rangle^b$ and $\rho(\lambda)^r=Z_m^{-1}\rho(\lambda)^b=Z_S\rho(\lambda)^b$. The relation $Z_mZ_S=1$ is satisfied automatically for the overlap fermion and, crucially, it avoids the systematic uncertainty from the additional chiral symmetry breaking in the Wilson-like actions. In terms of $Z_S^{-1}$, the residual RI/MOM scheme $Q^2$ dependence is proportional to $a^2$ and will vanish in the continuum limit, since the $a^2Q^2$ dependence shown in Fig.~\ref{fig:Zm} is roughly the same at all the lattice spacings we have. The statistical uncertainty in Fig.~\ref{fig:Zm} is at the 0.1\% level due to the volume source propagators~\cite{Chen:2017mzz} with different $Q^2$, but the total systematic uncertainties are $\sim$1.5\% which come predominantly from the estimated 4-loop effect in the perturbative matching between the RI/MOM and $\overline{\textrm{MS}}$ schemes ($\sim$90\% of this 1.5\%), and also from the value of $\Lambda_{\rm QCD}$, scale running, lattice spacing, and fit range, as shown in Tab.~\ref{tab:systematic}. Fortunately most of the systematic uncertainties are fully correlated at all the lattice spacings and are not enlarged in the continuum extrapolation. 

\begin{table*}[htbp]                   
\caption{{Error budget of $Z^{\overline{\textrm{MS}}}$(2~GeV) using the MOM scheme. All the values without unit are in percentage.}}  
\begin{tabular}{c|cccc}             
Lattice spacing  &0.0626(4) fm & 0.0711(3) fm & 0.0837(2) fm  & 0.1141(2) fm \\
\hline          
Statistics & 0.1 & 0.1 & 0.1 & 0.1 \\
\hline                                                    
Perturbative matching (missing 4-loop correction) & 1.2 &1.5 & 1.3 &1.5 \\
$\Lambda_{\rm QCD}$ (vary 5\%) & 0.2 & 0.3 & 0.3 & 0.3 \\
Scale running (4-loop v.s. 5-loop) & 0.1 & 0.1 & 0.1 & 0.1 \\
Lattice spacing (vary 1$\sigma$) & 0.3 & 0.2 & 0.1 & 0.1 \\
 Fit range (Enlarge the minimum $a^2Q^2$ by 1) & 0.1 & 0.1 & 0.1 & 0.1 \\
 \hline   
Total & 1.3 &1.5 & 1.3 &1.5 \\                                                                        
\end{tabular}  
\label{tab:systematic}                    
\end{table*}    

It is known that the matching coefficient between the symmetric momentum subtraction (RI/SMOM) scheme and the $\overline{\textrm{MS}}$ scheme is much closer to 1 up to the 3-loop level~\cite{Aoki:2007xm,Sturm:2009kb,Bednyakov:2020ugu,Kniehl:2020sgo}, so that the related systematic uncertainty can be significantly suppressed. However, in the following content of this section, we will argue in great detail that there are empirical ambiguities in fittings with the SMOM scheme, and it makes the total systematic uncertainty using the SMOM scheme even larger than that in the MOM case.

The $Z_S$ under the MOM and SMOM schemes are defined by 
\begin{align}\label{eq:smom}
&\frac{12Z^{\rm RI}_q(Q)}{\textrm{Tr}[\Lambda(p,p, I)]}_{p^2=-Q^2}=\frac{C_0}{m_q^2}+Z^{\rm MOM}_S(Q)+{\cal O}(m_q),\\
&\frac{12Z^{\rm RI'}_q(Q)}{\textrm{Tr}[\Lambda(p_1,p_2, I)]}_{p_1^2=p_2^2=q^2=-Q^2}=Z^{\rm SMOM}_S(Q^2)+{\cal O}(m_q),\\
& Z^{\rm RI}_q(Q)=\frac{Z_A}{48}\textrm{Tr}[\gamma_5\gamma_{\mu}\Lambda(p,p, \gamma_\mu\gamma_5)]_{p^2=-Q^2},\\
& Z^{\rm RI'}_q(Q)=\frac{Z_A}{48q^2}\textrm{Tr}[q_{\mu}\gamma_5q\!\!\!/\Lambda(p_1,p_2, \gamma_\mu\gamma_5)]_{p_1^2=p_2^2=q^2=-Q^2}\label{eq:ri_prime},
\end{align}
{where the normalization factor $Z_A$ and vertex function $\Lambda(p,p', \Gamma)$ are defined as}
\begin{align}
& Z_A=\left.\frac{2m_q\langle \bar{\psi}\gamma_5\psi|\pi\rangle}{m_{\pi}\langle\bar{\psi}\gamma_5\gamma_4\psi|\pi\rangle}\right|_{m_q\rightarrow 0},\\
 &\Lambda(p,p', \Gamma)= {\textrm{Tr}[\Gamma} S^{-1}(p_1)\sum_{x,y}e^{-i(p_1\cdot x-p_2\cdot y)}\nonumber\\
 &\quad\quad\quad\quad\ \langle \psi(x)\bar{\psi}(0){\Gamma}\psi(0)\bar{\psi}(y)\rangle S^{-1}(p_2)],
\end{align}
with $q=p_1-p_2$ and $S(p)=\sum_{x}e^{-i(p\cdot x)}\langle \psi(x)\bar{\psi}(0)\rangle$. Note that the quark self energy $Z_q$ is defined through the axial vector normalization constant $Z_A$, not the quark propagator directly. Ref.~\cite{Chang:2021vvx} has shown that $Z_q(Q)=\frac{1}{12p^2}\mathrm{Tr}[p\!\!\!/S^{-1}(p)]$ defined from the quark propagator has much larger discretization error compared to $Z^{\rm RI'}_q$ defined in Eq.~(\ref{eq:ri_prime}).
 
The matching factors of the scalar current in the MOM and SMOM cases with $n_f=3$ are ($a_s\equiv\alpha_s/\pi$):
\begin{align}
		\frac{Z_S^{\overline{\text{MS}}}}{Z_S^{\text{MOM}}}&=1+\sum_{i=1,2,3...} c^{\rm MOM}_{\rm N^iLO} a_s^i\nonumber\\
		&=1+1.333333a_s+9.93654a_s^2+84.403a_s^3\nonumber\\&+\mathcal{O}(a_s^4),\\
		\frac{Z_S^{\overline{\text{MS}}}}{Z_S^{\text{SMOM}}}&=1+\sum_{i=1,2,3...} c^{\rm SMOM}_{\rm N^iLO} a_s^i\nonumber\\
		&=1 + 0.16138  a_s + 0.686485 a_s^2 + 6.24424 a_s^3 \nonumber\\&+\mathcal{O}(a_s^4).	 
\end{align}
The 3-loop correction at $\mu=$4.0 GeV for the MOM case is 2.8\% and similar to the naive estimate from the power counting, $c^{\rm MOM}_{\rm N^3LO}\sim (c^{\rm MOM}_{\rm N^2LO})^2/c_{\rm NLO}=2.4$\%. Thus we estimate the uncertainty from missing the higher loop corrections as $c^{\rm MOM}_{\rm N^4LO}a_s^4\sim (c^{\rm MOM}_{\rm N^{ 3}LO})^2/c^{\rm MOM}_{\rm N^2LO}a_s^4=1.4$\%. In the SMOM case, the matching correction is much smaller but the correction at 3-loop level is $\sim 2$ times larger than the naive estimation, which suggests that the correction at higher loops can be larger than the naive guess $c^{\rm SMOM}_{\rm N^4LO}a_s^4\sim (c^{\rm SMOM}_{\rm N^{ 3}LO})^2/c^{\rm SMOM}_{\rm N^2LO}a_s^4=0.13$\% at $\mu=$ 4 GeV.

\begin{figure}[htbp]
    \centering
    \includegraphics[width=0.49\textwidth]{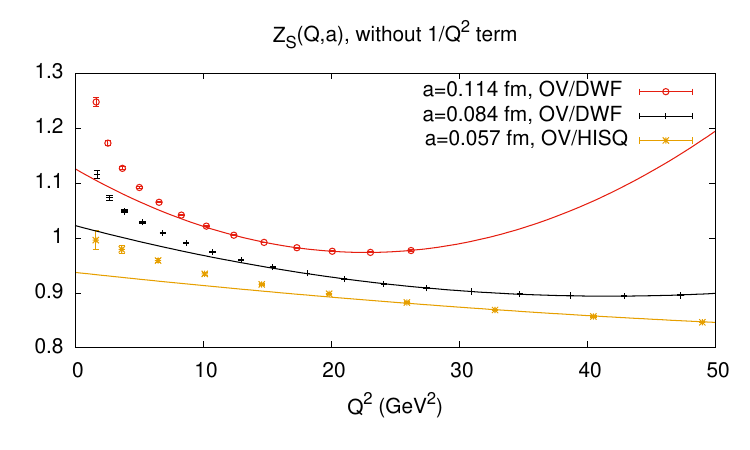}
     \includegraphics[width=0.49\textwidth]{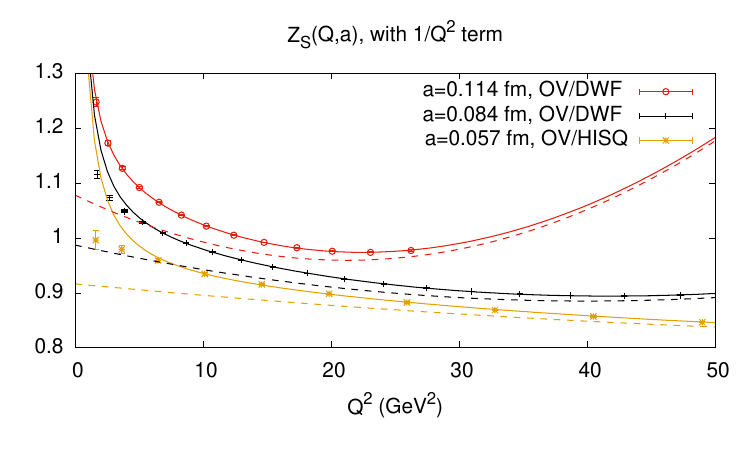}
    \caption{The scalar renormalization constant $Z_S$ at $\overline{\textrm{MS}}$ 2 GeV at three lattice spacings versus the SMOM scale $Q^2$. The lattice spacing dependence at $Q^2>20$ GeV$^2$ becomes larger with larger $Q$, implying an $a^2Q^2$ discretization error. {The solid curves in the {upper} panel show the fits using the polynomial form, and those in the {lower} panel use fits with an additional $1/Q^2$ term to describe the small-$Q^2$ behavior better. The dashed curves show the $Q^2$ dependence with the $1/Q^2$ term subtracted.} 
    \label{fig:mom0}}
\end{figure}   

\begin{table*}[t]                   
\caption{The parameters we obtained from two kinds of fits with or without the $c^{\rm SMOM}_{-1}/Q^2$ term. The lattice spacing dependence of the discretization error coefficients $c_{1,2}$ are mild.}  
\begin{tabular}{ccccccc}                                                                
$a$ (fm) & $|Q_{\rm min}|$ (GeV) &  $c^{\rm SMOM}_{-1}$ (GeV$^2$) & $Z_S$ & $c^{\rm SMOM}_{1}$ & $c^{\rm SMOM}_2$\\
\hline   
 0.114 fm& 3.2  & -- & 1.126(2) & -0.040(1) & 0.0026(1)\\   
  & 1.2  &0.32(1) & 1.078(3) & -0.033(1) & 0.0024(1)\\                           
\hline  
0.084 fm & 4.1  & -- & 1.023(2) & -0.034(1) & 0.0023(1)\\
  & 2.0  & 0.37(2) & 0.987(3) & -0.028(1) & 0.0020(1) \\
\hline   
0.057 fm  & 4.9  & -- & 0.937(3) & -0.030(1) & 0.0020(1)\\
  & 2.2  & 0.40(3) & 0.916(4) & -0.026(1) & 0.0018(1) \\
\end{tabular}  
\label{tab:fit}           
\end{table*}    

After the corresponding 2-loop matching between the SMOM and $\overline{\textrm{MS}}$ schemes, we obtain the $Z_S$ at $\overline{\textrm{MS}}$ 2 GeV using different SMOM scale $Q$, as shown in Fig.~\ref{fig:mom0}. Besides the physical point ensembles we used in the $\rho(\lambda)$ calculations, we also use the result using the overlap fermion on a MILC ensemble~\cite{MILC:2012znn} with the HISQ sea pion mass 310 MeV at 0.057 fm to show the lattice spacing dependence. As shown in Fig.~\ref{fig:mom0}, it is hard to find any linear region where the $a^2Q^2$ extrapolation can be made reliably as in the MOM case (see Fig.~\ref{fig:Zm}), as we found on the 48I ensemble with $a$=0.114 fm in the previous study~\cite{Bi:2017ybi}. At the same time, the $Q^2$ dependence below $Q^2<$ 20 GeV$^2$ is also non-linear. Thus we consider the following two empirical forms~\cite{Bi:2017ybi,Hasan:2019noy},
\begin{align}\label{eq:fit_zs}
Z^{\rm SMOM, a}_S(Q^2)&=\quad \quad\quad \quad\ \ \! Z_S+c^{\rm SMOM}_1a^2Q^2\nonumber\\&+c^{\rm SMOM}_2a^4Q^4,\\
Z^{\rm SMOM, b}_S(Q^2)&=\frac{c^{\rm SMOM}_{-1}}{Q^2}+Z_S+c^{\rm SMOM}_1a^2Q^2\nonumber\\&+c^{\rm SMOM}_2a^4Q^4,
\end{align}
and tune the minimum $Q^2$ used in each fit to make the $\chi^2$/d.o.f. smaller than 1. The results we obtained are summarized in Tab.~\ref{tab:fit} and illustrated in Fig.~\ref{fig:mom0}. We can see that generally the form with the $1/Q^2$ term can have a better description of the data at small $Q^2$, but the $c^{\rm SMOM}_{1,2}$ from two fits at different lattice spacings are close to each other. At the same time, the coefficient $c^{\rm SMOM}_{-1}$ does not vanish in the continuum limit, even though the non-linear behavior at small $Q^2$ seems to be milder at smaller lattice spacings. The difference between $Z_S$ from the two fits at the three lattice spacings are 0.048(2), 0.036(2) and 0.021(2) respectively, which decreases with $a$ and thus is likely to be an additional discretization error. Thus we will take the $Z_S$ from the polynomial fit as the central value and the difference of $Z_S$ from two fits as a systematic uncertainty.

\begin{figure}[htbp]
    \centering
    \includegraphics[width=0.49\textwidth]{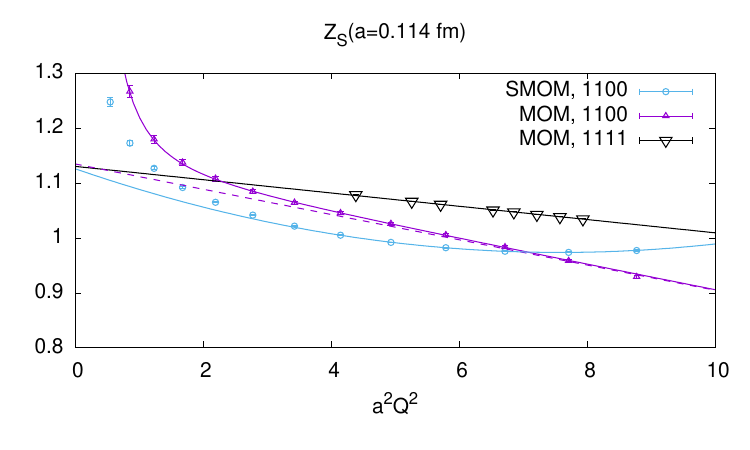}
    \includegraphics[width=0.49\textwidth]{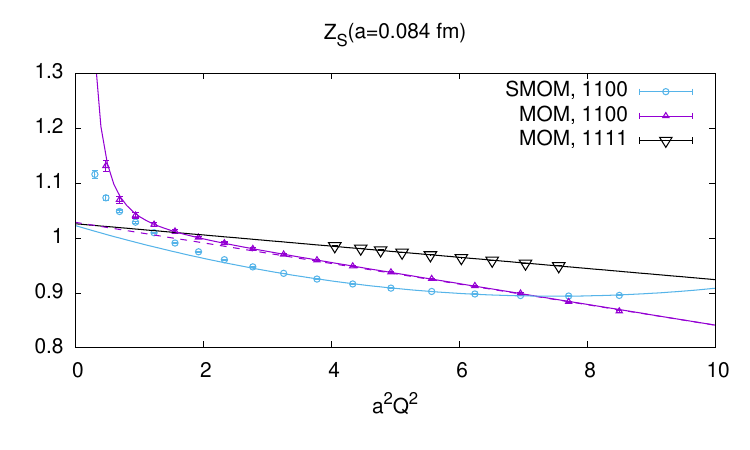}
    \caption{The scalar renormalization constant $Z_S$ at $\overline{\textrm{MS}}$ 2 GeV and two lattice spacings, using two kinds of of regularization independent schemes.  That using the SMOM scheme with momenta $p=(k,k,0,0)$ and $p'=(k,0,k,0)$ (blue dots) has much larger discretization error than those with the MOM scheme using either $p=(k,k,0,0)$ (purple triangles) or the body-diagonal momenta (black triangles), and the non-perturbative effects exist in both the SMOM and MOM cases at small $Q^2$.
    \label{fig:mom}}
\end{figure}

Then we can compare the values obtained using the SMOM scheme with those using the MOM scheme. As seen in Fig.~\ref{fig:mom}, the extrapolated values using the SMOM scheme (the intercepts of the blue curves) {$Z_S(a=0.114~ \textrm{fm})=1.126(5)(48)$ and $Z_S(a=0.084~ \textrm{fm})=1.023(3)(36)$ at $\overline{\textrm{MS}}$ 2 GeV}, are consistent with the values $Z_S(a=0.114~ \textrm{fm})=1.117(1)(17)$ and $Z_S(a=0.084~ \textrm{fm})=1.028(1)(13)$ using the MOM scheme with the body-diagonal momenta (the intercepts of the black lines) and 3-loop matching. We also calculated the $Z_S$ using the MOM scheme but with momenta $p=(k,k,0,0)$ with different $k$ (purple triangles). With an empirical form $Z^{\rm MOM}(Q^2)=\frac{C^{\rm MOM}_0}{Q^4}+Z_0+C^{\rm MOM}_1a^2Q^2$, which can describe the data in the entire range $a^2Q^2\in[0.5,9.0]$ with $\chi^2$/d.o.f.$\sim$ 1, the extrapolated values (the intercepts of the purple dashed curves which correspond to the $Q^2$ dependence with the $1/Q^4$ term subtracted) are also consistent with those from the other two cases.

Thus even though the perturbative convergence of the matching coefficient between the SMOM and $\overline{\textrm{MS}}$ schemes is much better than the MOM case up to 3-loop level, the $a^2Q^2$ extrapolation with non-linear $Q^2$ terms can be quite sensitive to the empirical form used and thus introduces additional systematic uncertainty. If one trivially assumes a good perturbative matching convergence, and then uses the value at $Q=2$ GeV directly or does the linear $a^2Q^2$ extrapolation at small $a^2Q^2$, the corresponding $Z_S$ and also $\Sigma$ can be 5--10\% larger. It is also reported in Ref.~\cite{Hasan:2019noy} that using the MOM or SMOM scheme can introduce a systematic uncertainty on the scalar current at the 10\% level for the clover fermion.

Our results of the scalar current renormalization constants are listed in Tab.~\ref{tab:ensemble}, with two uncertainties from the statistics and systematics. Note that the systematic uncertainty from the perturbative matching will not be enlarged during the continuum extrapolation as they are correlated at different lattice spacings.

\begin{figure}[htbp]
	\centering
	\includegraphics[width=0.48\textwidth]{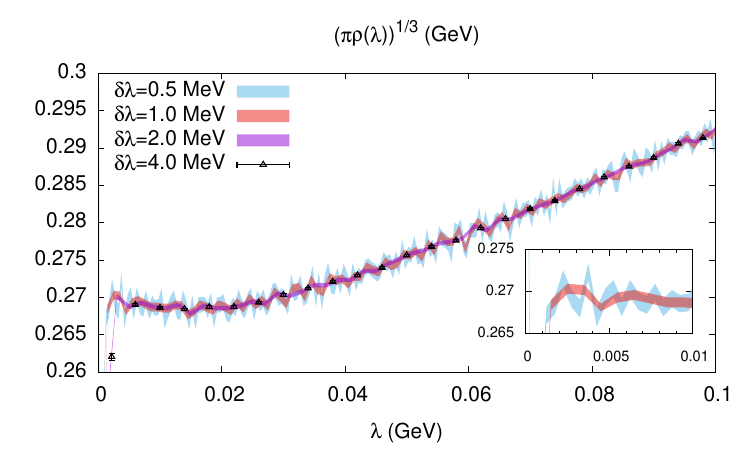}
	\caption{The $\lambda$ dependence of $\big(\pi\rho(\lambda)\big)^{1/3}$ at 0.084 fm. Different {bands} are for the results with different bin sizes. The number of eigenvalues in the bin will be smaller than 25 when $\delta \lambda<$ 3 MeV and causes obvious fluctuations, but the data are still flat (up to the statistical fluctuation) when $\lambda\gtrsim$ 1 MeV even with the smallest bin size.}
	\label{fig:density}
\end{figure}

\section{Chiral Condensate}

After the above detailed discussion of renormalization,
we now switch to the numerical results 
of the overlap Dirac spectrum and the determination of the chiral condensate. In Fig.~\ref{fig:density}, we plot our $\big(\pi\rho(\lambda)\big)^{1/3}$ results at $a=0.084$ fm with 0.5, 1, 2, and 4 MeV bin size, respectively. 
The uncertainty and fluctuation of our results become much larger with smaller bin size since the number of eigenvalues in each bin gets fewer and violates the central limit theorem requirement. However, the uncertainty is just of 1\% level with the smallest bin size. 
And with this bin size, only the first two data points drop significantly, which suggests that the finite volume effect is relatively small with the 5.5 fm box at the physical pion mass. 
Based on the standard statistical requirement of having at least 25 samples in each bin ($\sim$ 3 MeV bin size), we choose to use a 3.5 MeV bin size in the remaining numerical analysis of this work. 
Using such a bin size, our statistical uncertainty is an order of magnitude smaller than those of all the previous $\rho(\lambda)$ calculations~\cite{Giusti:2008vb,Fukaya:2009fh,Fukaya:2010na,Cichy:2013gja,Engel:2014eea,Cossu:2016eqs}. This is understandable since the number of eigenvalues is almost proportional to the physical volume, and we used the largest volume to date in this work.

\begin{figure}[htbp]
    \centering
    \includegraphics[page=1, width=0.49\textwidth]{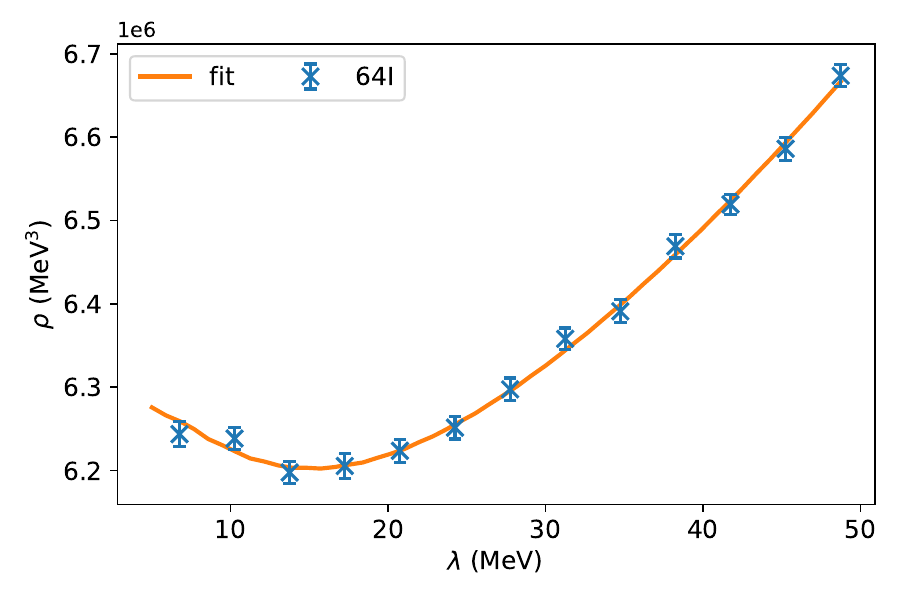}
    \caption{A fitting example of 
    the spectral density using the $\rm SU(3)$ chiral form on the 64I lattice.}
    \label{Fig:fit_exp}
\end{figure}

We use the NLO chiral expression Eq.~(\ref{eq:def})
to fit for the chiral condensate. 
Since Eq.~(\ref{eq:def}) is a general form for
arbitrary number of non-degenerate quarks~\cite{Damgaard:2008zs},
we can use it 
in both the $N_f=2+1$ and $N_f=2$ cases to determine
the $\rm SU(3)$ and $\rm SU(2)$ chiral condensates respectively.
The $N_f=2$ form is valid for 
$\lambda$ much smaller than the strange quark mass
due to the fact that
the third flavor (the strange quark) mainly accounts for the rising behavior of $\rho(\lambda)$ at large $\lambda$.
We use three lattices 48I, 64I and 48IF to
control the lattice spacing dependence.
Numerically, in these fittings,
only the pion decay constants (denoted by $F_0$ for the $\rm SU(3)$
case and $F$ for the $\rm SU(2)$ case)
and the chiral condensates (denoted by $\Sigma_0$ for the $\rm SU(3)$ case
and $\Sigma$ for the $\rm SU(2)$ case)
are set to be free parameters while
the pion/Kaon masses determined from each lattice 
as collected in Tab.~\ref{tab:ensemble} are used as inputs. 
An example of fitting in the $\rm SU(3)$ case on the 64I lattice is shown in Fig.~\ref{Fig:fit_exp}
and the corresponding $\chi^2/d.o.f.$ is 0.35 \red{with the correlation among the different data points included. All the ChPT fits performed in this work are correlated fits, as the data points exhibit strong correlations with those in neighboring bins.}

\begin{figure}[!h]
    \centering
    \includegraphics[page=1, width=0.49\textwidth]{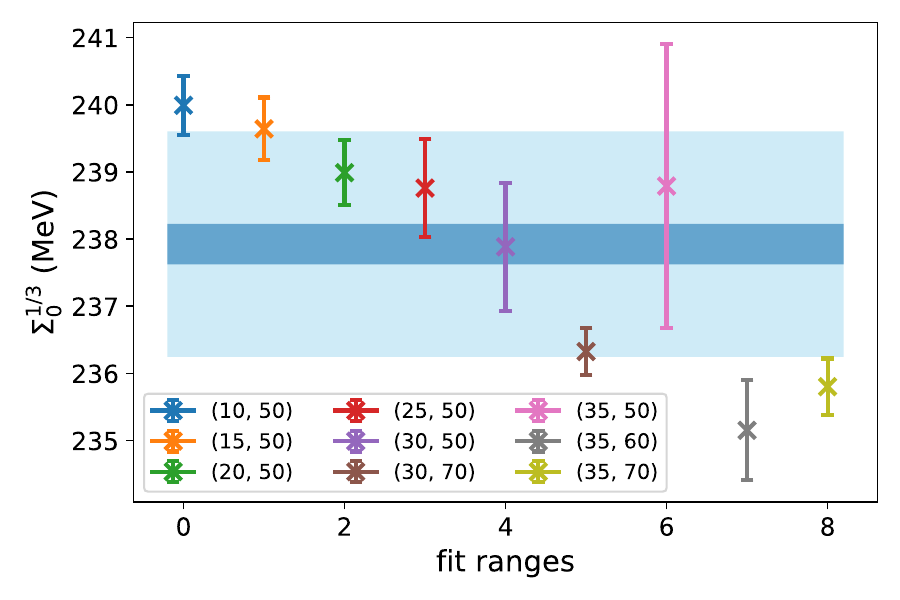}
    \includegraphics[page=1, width=0.49\textwidth]{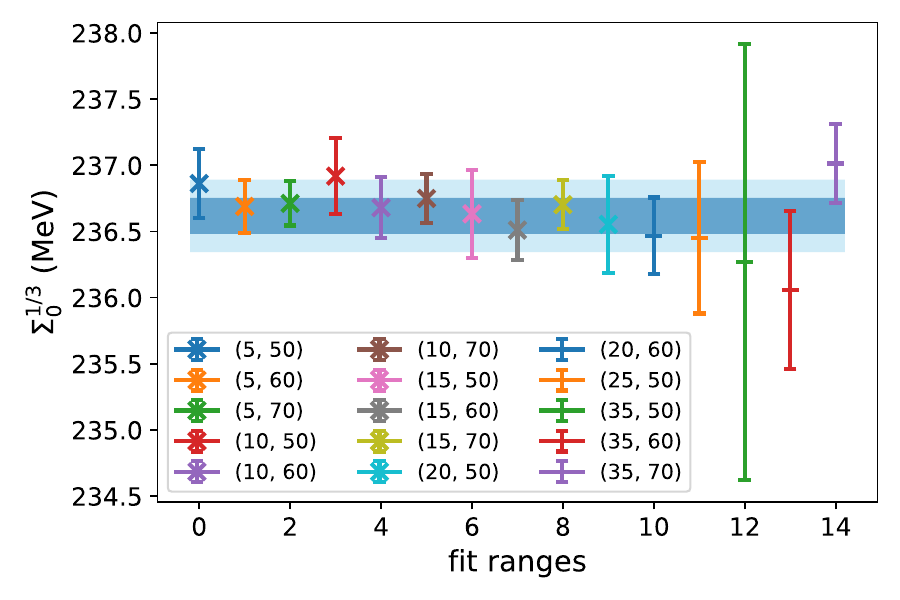}
    \includegraphics[page=1, width=0.49\textwidth]{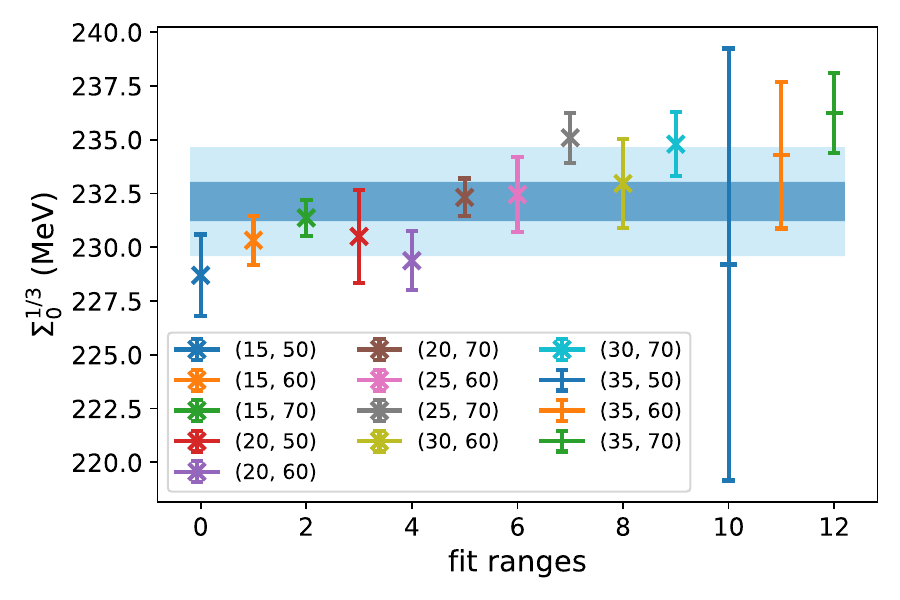}
    \caption{The fitting results of $\Sigma_0$ with different fit ranges.
    The starting and ending points of the fit ranges are listed in the legend.
    The three panels are for the 48I, 64I and 48IF lattices, respectively. The narrow blue band in each panel represents the statistical error of
    a constant fit to all the data points, and the wider light blue band is to show the 
    total error with the systematic error caused by 
    fit ranges added in quadrature.}
    \label{Fig:fit_SU3_con}
\end{figure}

\begin{figure}[!h]
    \centering
    \includegraphics[page=1, width=0.49\textwidth]{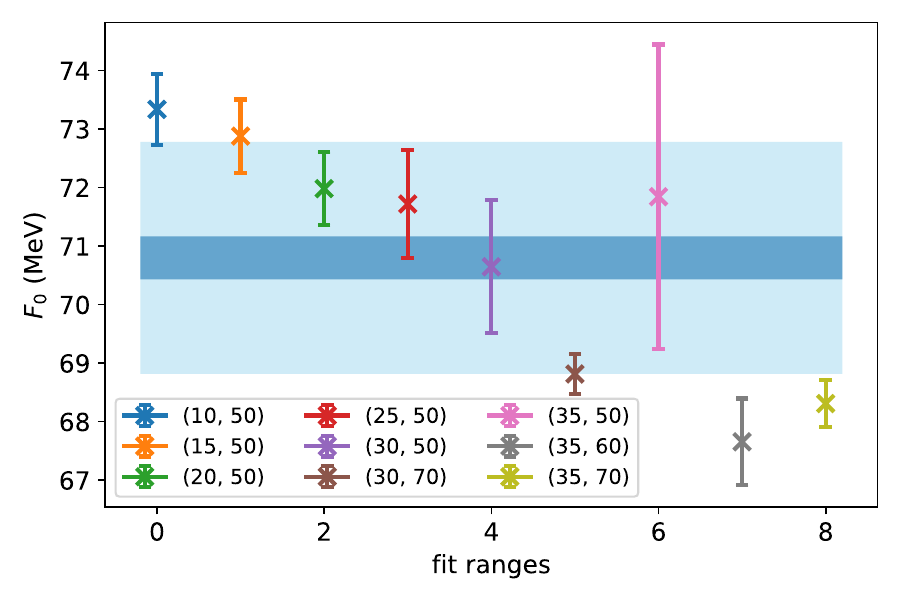}
    \includegraphics[page=1, width=0.49\textwidth]{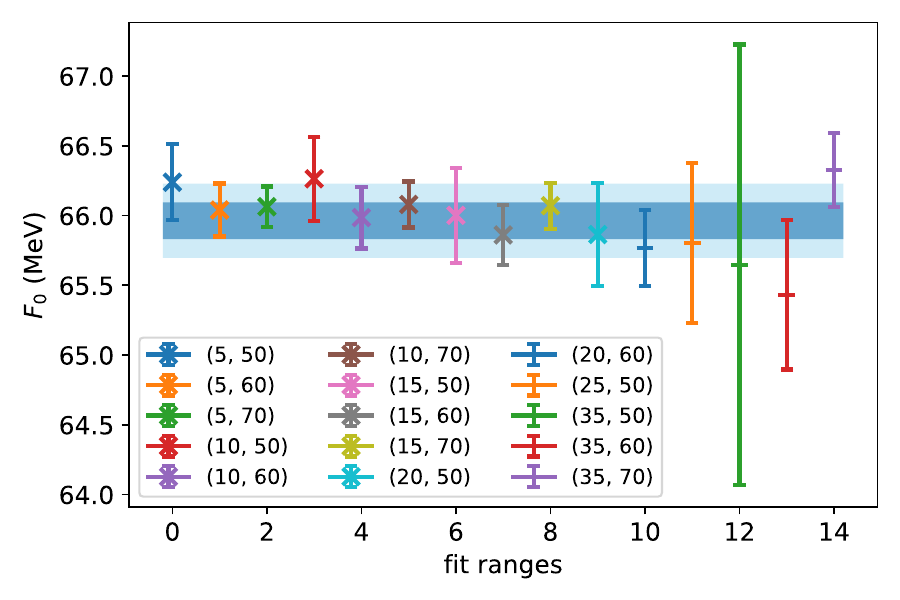}
    \includegraphics[page=1, width=0.49\textwidth]{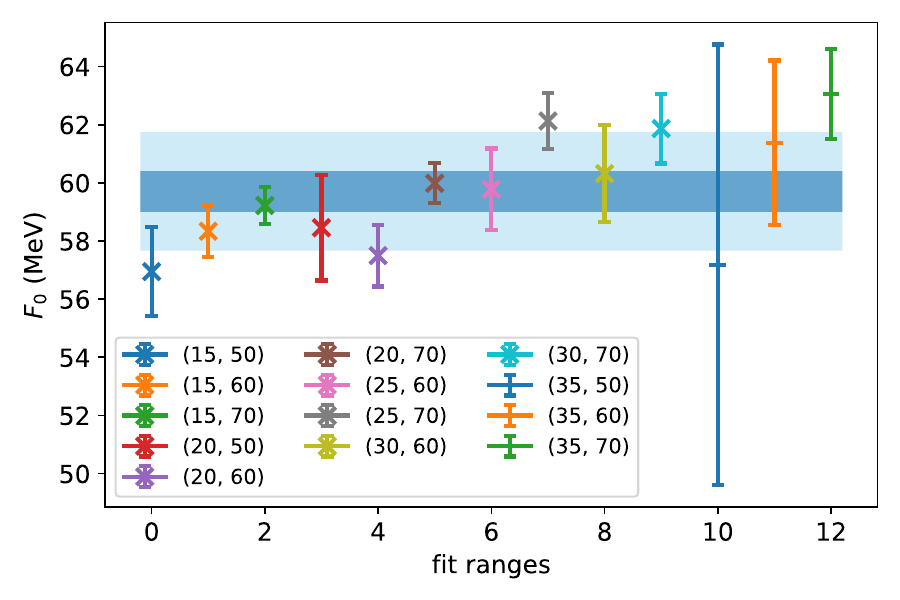}
    \caption{The same as Fig.~\ref{Fig:fit_SU3_con} but for $F_0$.} 
    \label{Fig:fit_SU3_F}
\end{figure}

\begin{figure}[!h]
    \centering
    \includegraphics[page=1, width=0.49\textwidth]{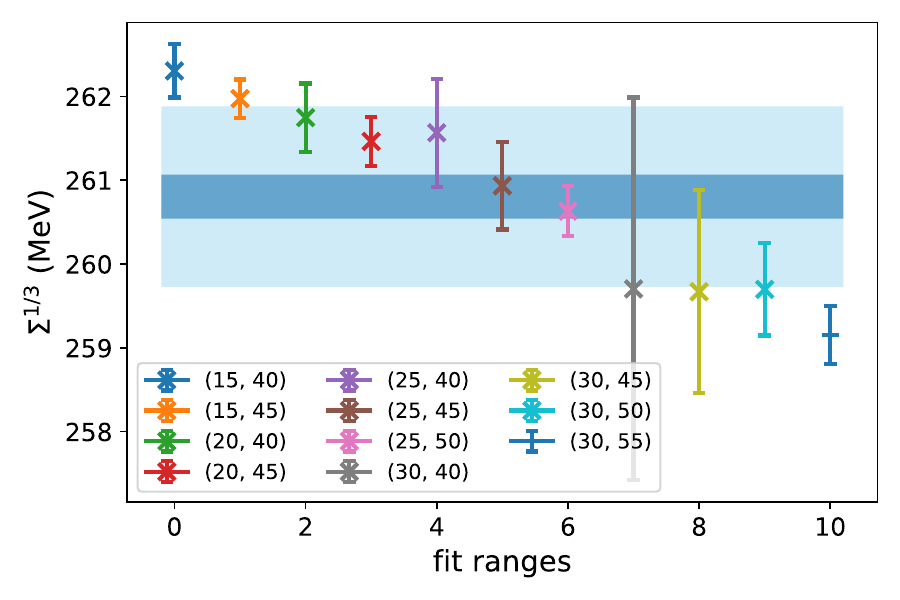}
    \includegraphics[page=1, width=0.49\textwidth]{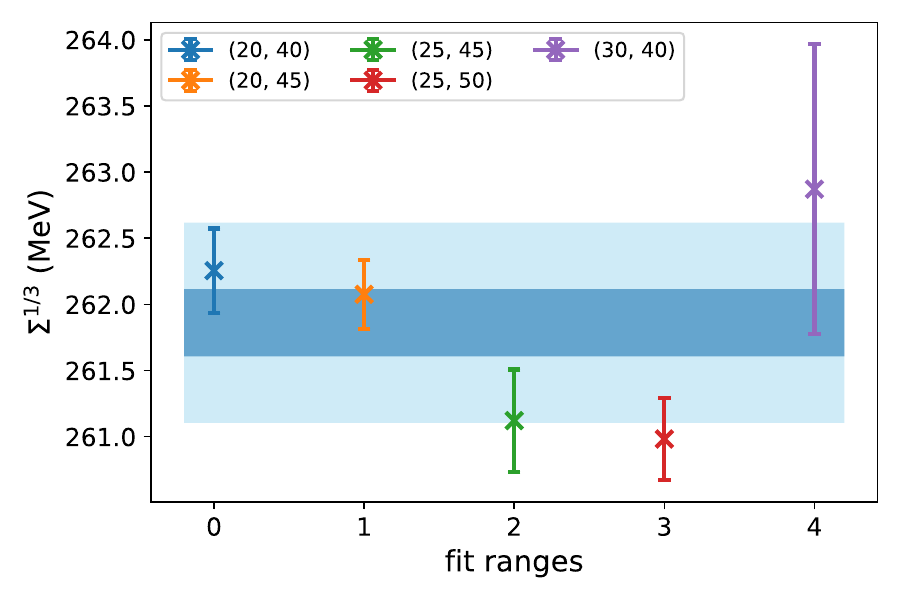}
    \includegraphics[page=1, width=0.49\textwidth]{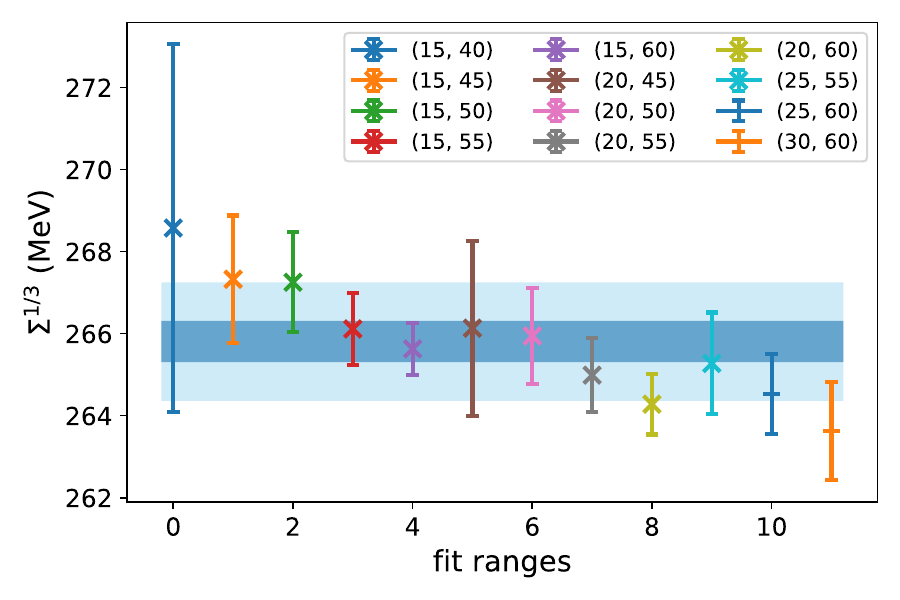}
    \caption{The same as Fig.~\ref{Fig:fit_SU3_con} but for $\Sigma$.}
    \label{Fig:fit_SU2_con}
\end{figure}

\begin{figure}[!h]
    \centering
    \includegraphics[page=1, width=0.49\textwidth]{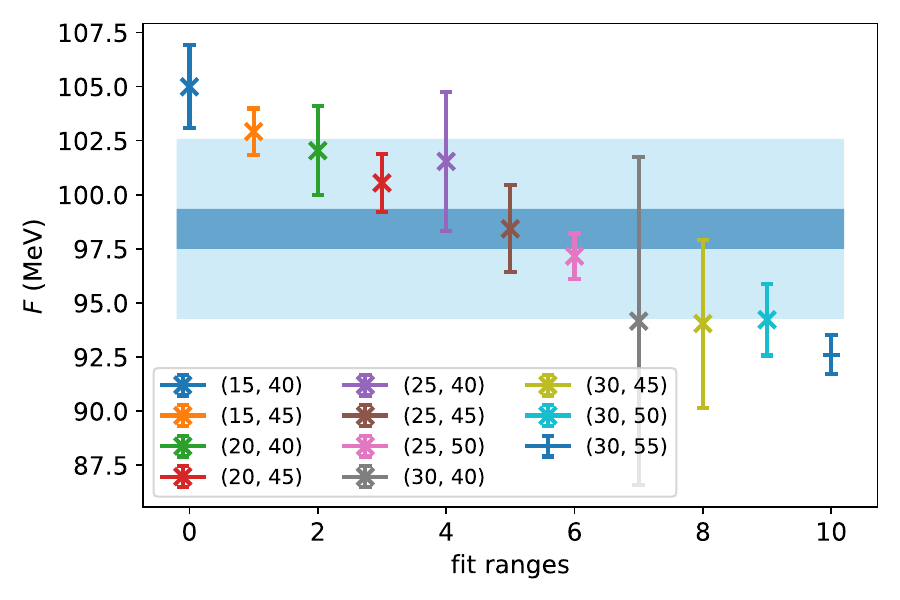}
    \includegraphics[page=1, width=0.49\textwidth]{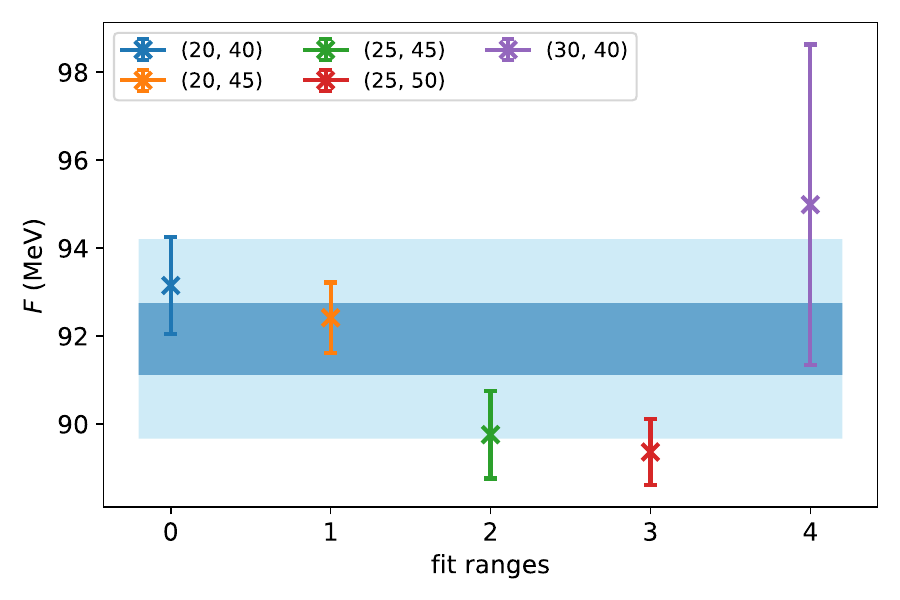}
    \includegraphics[page=1, width=0.49\textwidth]{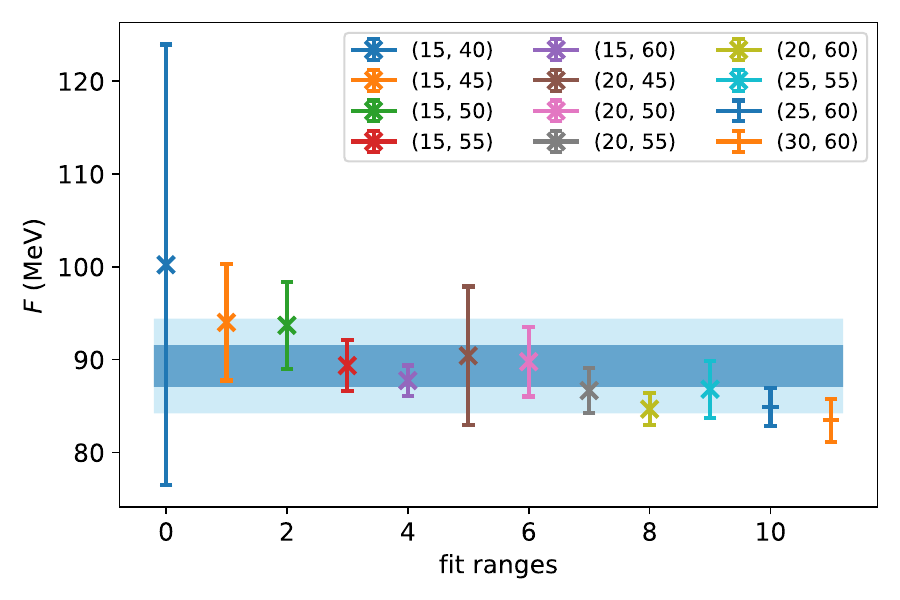}
    \caption{The same as Fig.~\ref{Fig:fit_SU3_con} but for $F$.}
    \label{Fig:fit_SU2_F}
\end{figure}

To well control the systematic uncertainties caused by
using different fitting ranges, the fittings are
carried out several times with different fitting ranges. Specifically, 
the starting $\lambda$ is from 5 to 35 MeV with a
5 MeV step and the ending point is from 50 to 70 MeV
with a 10 MeV step for the $\rm SU(3)$ case,
while the starting $\lambda$ is from 15 to 30 MeV with a
5 MeV step and the ending point is from 40 to 60 MeV
with also a 5 MeV step for the $\rm SU(2)$ case,
respectively. The fitting ranges are thus chosen to 
cover most ranges that lead to reasonable $\chi^2$, 
and all the fittings results with $\chi^2/d.o.f.<1.5$
are plotted in Fig.~\ref{Fig:fit_SU3_con},
Fig.~\ref{Fig:fit_SU3_F},
Fig.~\ref{Fig:fit_SU2_con} and
Fig.~\ref{Fig:fit_SU2_F}
for $\Sigma_0$, $F_0$, $\Sigma$ and $F$
respectively.
The three lattices are in different panels in each figure, and in each panel
the narrow blue band represents the statistical error of
a constant fit to all the data points.
The systematic errors of fit ranges are estimated by calculating the standard deviation of the data points, and the total uncertainties
with the systematic errors added in quadrature
are denoted by the wider light blue bands.
From those figures one can conclude three points:
a) The decay constant and the chiral condensate are strongly
correlated in the fittings (the effects of changing fitting ranges on the
condensate and decay constant are in a very similar way),
which is understandable since in the chiral and
infinite volume limit
\begin{equation}
    \rho(\lambda)\sim\Sigma\left[1+\left(\frac{N_f^2-4}{N_f}\right)\frac{\Sigma}{32\pi F^4}\lambda\right],
\end{equation}
such that the chiral condensate is determined mainly from
$\rho(\lambda)$ at small $\lambda$ region and then the
decay constant is determined in the form of a 
ratio $\frac{\Sigma}{F^4}$ that controls the $\lambda$ dependence.
b) The errors of the 48IF lattice 
are much larger than those of the other two lattices because
the 48IF lattice has the smallest physical volume and
thus the smallest number of eigenvalues.
And c) in most cases,
the fitting systematic uncertainties are larger
than the statistical ones, which is partially due to
the the high statistical precision of our
lattice data. We will keep both the errors
and use the total uncertainties in the following further analysis.

\begin{figure}[!h]
    \centering
    \includegraphics[page=1, width=0.49\textwidth]{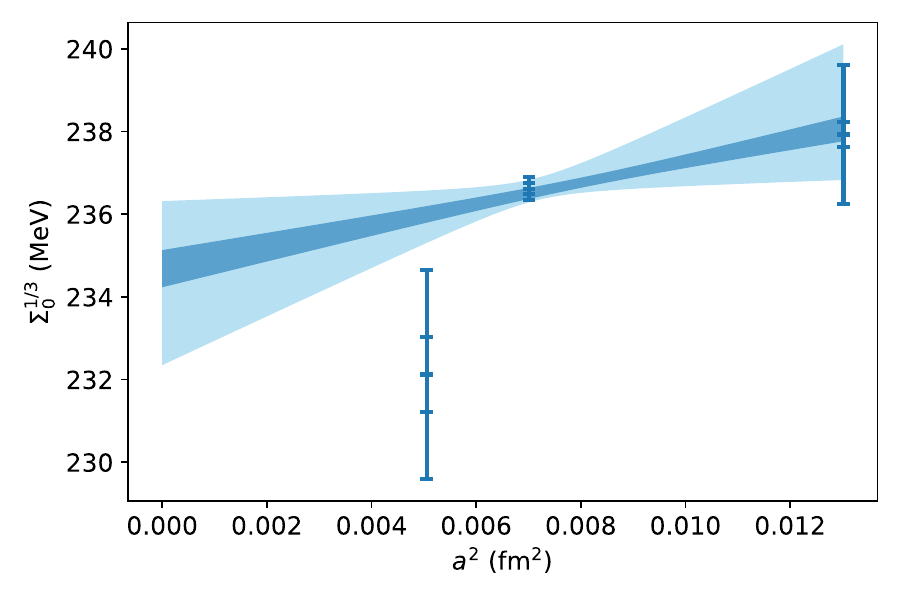}
    \includegraphics[page=1, width=0.49\textwidth]{eigen_values_SU3_continuum_1.pdf}
    \caption{The continuum extrapolation for the $\rm SU(3)$
    case. The two panels are for $\Sigma_0$ and $F_0$ respectively. Similar to Fig.~\ref{Fig:fit_SU3_con},
    the two bands and the two errors of each data point
    denote the statistical error and the
    total error with the systematic error caused by 
    fit ranges added in quadrature.}
    \label{Fig:continue:SU3}
\end{figure}

\begin{figure}[!h]
    \centering
    \includegraphics[page=1, width=0.49\textwidth]{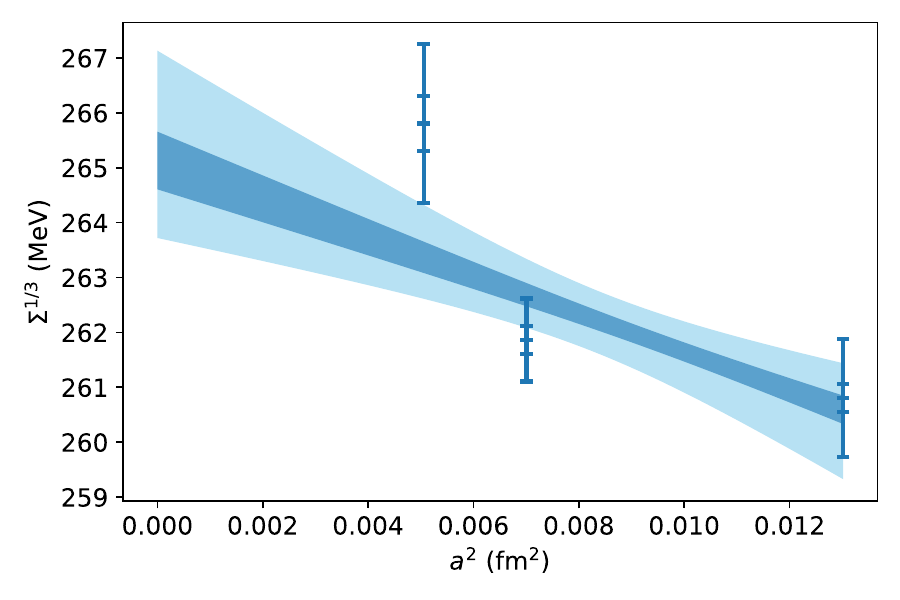}
    \includegraphics[page=1, width=0.49\textwidth]{eigen_values_SU2_continuum_1.pdf}
    \caption{The same as Fig.~\ref{Fig:continue:SU3}
    but for the $\rm SU(2)$ case.}
    \label{Fig:continue:SU2}
\end{figure}

After all the fittings of $\rho(\lambda)$ are done on each lattice, 
a lattice spacing extrapolation is performed using a linear form in $a^2$
to push the results to the continuum limit
as shown in
Fig.~\ref{Fig:continue:SU3} and Fig.~\ref{Fig:continue:SU2}.
Again, the two bands and the two errors of each data point
denote the statistical error and the total error with the 
systematic error caused by fit ranges added in quadrature.
One can see that
with the total error taken into consideration,
the linear fits work fine.
The results of the 48IF lattice drift to some extent
from the linear behavior, which is possibly due to
statistical fluctuation and the fact that this is a
lattice with relatively heavy $u$ and $d$ quark masses.
We take the difference between
the extrapolated results
and the central values 
of the 48IF lattice to be the
systematic error of the 
continuum extrapolation
and possible pion mass effects.

\begin{table}[htbp]
    \centering
        \caption{The fitting results on the 64I lattice with different low-energy constant $L_6$.}
    \begin{tabular}{c|c|c}
         $L_6$ $\rm SU(3)$   & $0.01\times10^{-3}$ & $0.16\times10^{-3}$  \\
         \hline
      $F_0$ (MeV) &   66.24(27) & 55.71(32) \\
      $\Sigma_0^{1/3}$ (MeV)  & 236.86(26) & 211.04(47) \\
      \hline
       \hline\\[-1em]
       $L_6$  $\rm SU(2)$   & $-0.23\times10^{-3}$ & $0.101\times10^{-3}$  \\
       \hline
      $F$ (MeV) &   92.41(80) & 91.30(81) \\
      $\Sigma^{1/3}$ (MeV)  & 262.08(26) & 259.96(30) 
    \end{tabular}

    \label{tab:l6}
\end{table}

\begin{table}[htbp] 
    \centering
        \caption{The complete error budget
        of the chiral condensates and decay constants. 
        All the values are converted in percentage.}
    \begin{tabular}{c|c|c|c|c}
       & $F_0$ & $\Sigma_0^{1/3}$ & $F$ & $\Sigma^{1/3}$ \\
       \hline
       statistical & 0.82 & 0.19 & 2.3 & 0.20\\
       \hline
       fit ranges  & 3.8 & 0.85 & 7.0 & 0.66\\
       lattice spacing & 1.9 & 0.93 & 6.2 & 0.15\\
       NLO chiral form & 16 & 11 & 1.5 & 0.81\\
       renormalization & 1.1 & 1.1 & 1.1 & 1.1 \\
       scale setting & 0.30 & 0.30 & 0.30 & 0.30 \\
        \hline
       total systematic & 17 & 11 & 9.5 & 1.6
    \end{tabular}
    \label{tab:error}
\end{table}

In addition to the fitting ranges and continuum extrapolation,
the NLO low-energy constant $L_6$ used in the
chiral form can also cause systematic uncertainties.
In the original paper~\cite{Damgaard:2008zs},
they use the value $L_6=0.05\times10^{-3}$
in both the $\rm SU(3)$ and the $\rm SU(2)$ cases.
The FLAG review~\cite{FlavourLatticeAveragingGroupFLAG:2021npn} 
collects two values of $\rm SU(3)$ $L_6$
and we use the $N_f = 2 + 1$ one $L_6$ = $0.01(34)\times10^{-3}$~\cite{MILC:2010hzw}
to get the our central values
and the $N_f = 2 + 1+1$ one $L_6$ = $0.16(20)\times10^{-3}$~\cite{Dowdall:2013rya}
to get a second set of results;
the uncertainties are estimated 
simply to be the differences.
For the $\rm SU(2)$ case,
FLAG does not collect any results, and
we use the values from ref.~\cite{Boyle:2015exm}
where the partially quenched $\chi$PT is adopted.
In this case,
we use the NNLO value with 450 MeV cut $L_6=-0.23\times10^{-3}$
to get the central values in our fittings and
the NLO one with 450 MeV cut $L_6=0.101\times10^{-3}$
to estimate the uncertainty.
The fitting results on the 64I lattice with different $L_6$'s
are listed in Tab.~\ref{tab:l6} for a clear demonstration
on the effects of changing $L_6$.
This uncertainty can also be treated as part of
the uncertainty of using the NLO chiral form.
The NNLO effects of calculating $\Sigma_0$ and $F_0$ using $\chi$PT form on 2+1-flavor lattice data with physical strange quark mass
can also be checked as $\frac{m_K^4}{(4\pi F_0)^4}\sim10\%$, which is weaker than the $L_6$ effects.
Thus, we use different $L_6$ values to estimate the systematic uncertainties of using only the NLO chiral form.

We have also considered the systematic uncertainties from the 
determination of renormalization constants and lattice scale settings.
A complete error budget is listed in Tab.~\ref{tab:error}.
It shows that the greatest uncertainty in the $\rm SU(3)$
case comes from the low-energy constant $L_6$ of
the chiral form.
Actually, to our best knowledge, our study is so far the only one that uses the complete NLO chiral form and takes the $L_6$ effects into account.
In the $\rm SU(2)$ case,
the fit range and continuum extrapolation also play an important role.
This can be understood since the $\rm SU(2)$ case
is more sensitive to the fit range used, as discussed above.
Our final prediction of the chiral condensates and pion
decay constants in both the $\rm SU(3)$
and $\rm SU(2)$ cases are
\begin{equation}
    F_0 = 58.6(0.5)(10.0)~{\rm{MeV}},
\end{equation}
\begin{equation}
    \Sigma_0^{1/3} = 234.3(0.5)(25.8)~{\rm{MeV}},
\end{equation}
\begin{equation}
    F = 84.1(1.9)(8.0)~{\rm{MeV}},
\end{equation}
and
\begin{equation}
    \Sigma^{1/3} = 265.4(0.5)(4.2)~{\rm{MeV}},
\end{equation}
where the two errors are
the statistical one and the total systematic one respectively. \red{Note that the convergence of SU(3) ChPT can be poor with physical quark masses. Therefore, our predictions here may suffer from uncontrollable systematic uncertainties unless a gauge ensemble with significantly lighter strange quark masses is available.}

\begin{figure*}[htbp]
    \centering
    \includegraphics[page=1, width=0.49\textwidth]{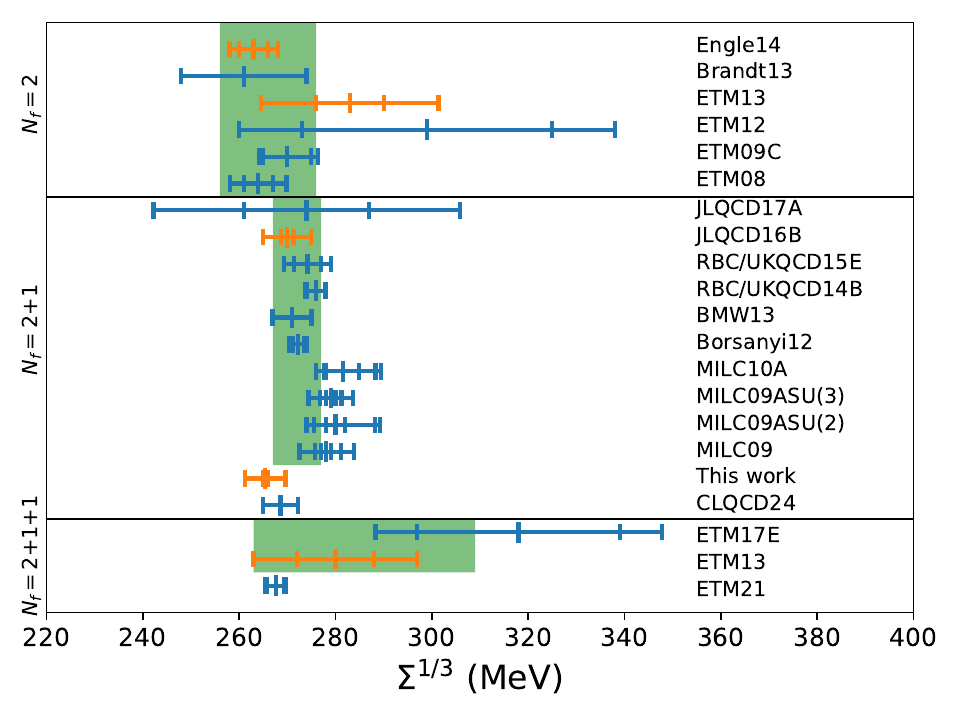}\\
    \includegraphics[page=1, width=0.49\textwidth]{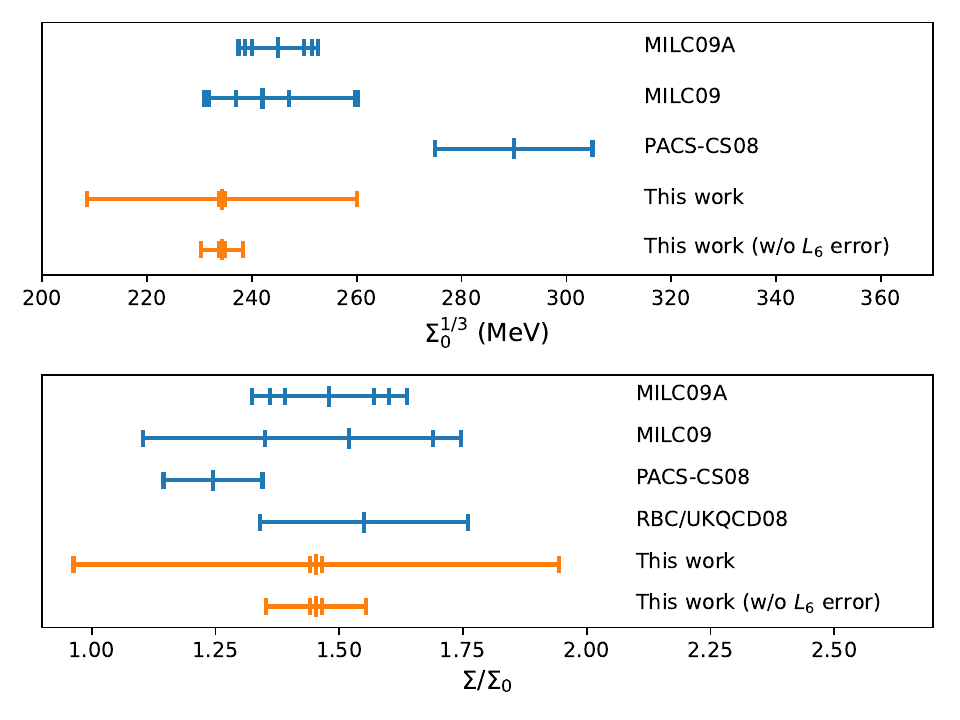}
    \includegraphics[page=1, width=0.49\textwidth]{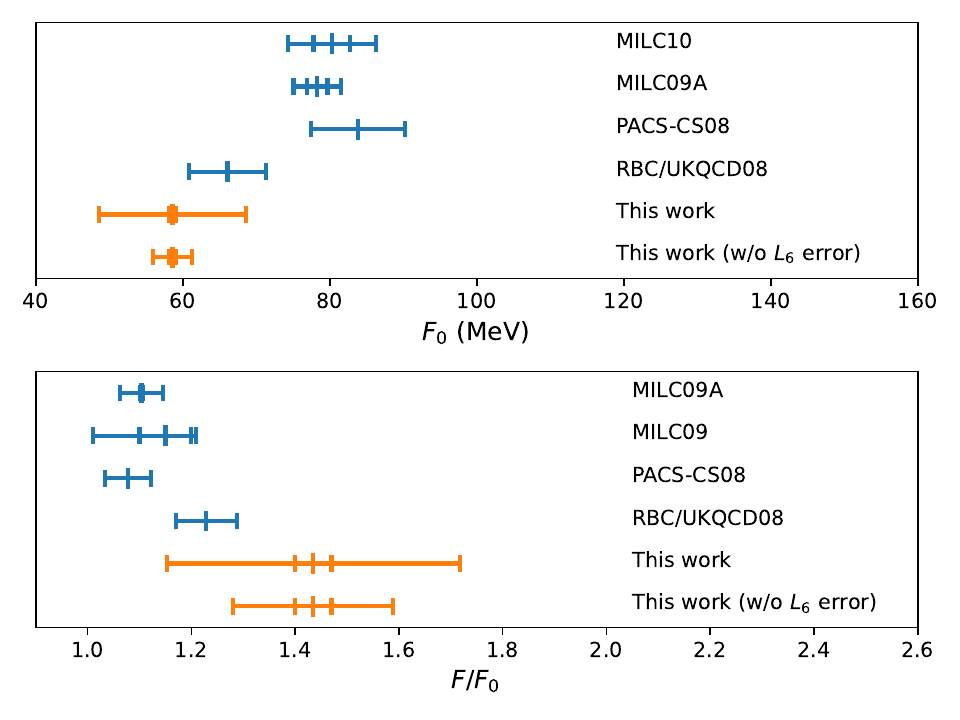}
    \caption{Results of $\rm SU(2)$ chiral condensate $\Sigma^{1/3}$ (top panel), $\rm SU(3)$ chiral condensate $\Sigma_0^{1/3}$ and the ratio of $\Sigma/\Sigma_0$ (lower left panel), 
    and $\rm SU(3)$ decay constant $F_0$ and the ratio of $F/F_0$ (lower right panel) from different lattice calculations. Numbers are taken from Refs.~\cite{Engel:2014eea}(Engle14),
    \cite{Brandt:2013dua} (Brandt13),
    \cite{Burger:2012ti} (ETM12),
    \cite{Baron:2009wt} (ETM09C),
    \cite{Frezzotti:2008dr} (ETM08),
    \cite{Aoki:2017paw} (JLQCD17A),
    \cite{Cossu:2016eqs} (JLQCD16B),
    \cite{Boyle:2015exm} (RBC/UKQCD15E),
    \cite{Blum:2014tka} (RBC/UKQCD14B),
    \cite{Durr:2013goa} (BMW13),
    \cite{Borsanyi:2012zv} (Borsanyi12),
    \cite{Bazavov:2010yq} (MILC10A),
    \cite{Bazavov:2009fk} (MILC09ASU(3), MILC09ASU(2)), 
    \cite{Bazavov:2009bb} (MILC09),
    \cite{Alexandrou:2017bzk} (ETM17E),
    \cite{Cichy:2013gja} (ETM13),
    \cite{ExtendedTwistedMass:2021gbo} (ETM21),
    \cite{Aoki:2008sm} (PACS-CS08),
    \cite{Allton:2008pn} (RBC/UKQCD08),
    and
    \cite{CLQCD:2023sdb} (CLQCD2024).
    In the top panel, orange points with error bars are from lattice works using the Dirac spectrum method, 
    while blue ones are from lattice works using other methods. The green bands in the top panel indicate the corresponding lattice average values from FLAG21~\cite{FlavourLatticeAveragingGroupFLAG:2021npn}.
    Points that are not covered by the bands are new lattice results which have not been included in the FLAG average yet. 
    \label{fig:com}}
\end{figure*}

Compared to all the previous determinations shown in Fig.~\ref{fig:com},
we have reasonably good results for $\Sigma$.
Actually, its statistical uncertainty is very small.
For $F_0$ and $\Sigma_0$,
due to the large uncertainties from the changing of $L_6$,
we have relatively large total errors. However, 
as we have emphasized before, this is method-related
and this is the first time the $L_6$ NLO uncertainty is carefully checked.
Thus, in this sense, this study provides
the currently best results using the Dirac spectrum method.
We have also determined the ratio $\Sigma/\Sigma_0= 1.45(1)(49)$
and 
$F/F_0= 1.44(4)(28)$,
and they reflect the difference between the $N_f=$ 2 chiral limit ($m_s\sim$ 90 MeV) and the $N_f=$ 3 chiral limit ($m_s=$ 0), as in the NLO expression of $Z_v$.
We also add the results without the $L_6$ error in the plots for comparison.

\section{Other Sea Information}

\begin{table*}
\caption{Summary of the low-energy constants and sea quark masses obtained from the fit using the $\rm SU(3)$ finite volume NLO PQ$\chi$PT forms in Ref.~\cite{Damgaard:2008zs}. All the values except $\chi^2$ are in unit of MeV.}  

\begin{tabular}{c|llllc}                                  

\text{Lattice spacing }&  $\Sigma_0^{1/3}$ & $F_0$ & $m_l$ & $m_s$ & $\chi^2$/d.o.f.   \\
\hline   
0.114 fm  & 224.9(1.5) & 57.3(1.4) & 2.74(06) & 87.0(5.1) & 0.99 \\

0.084 fm & 233.6(1.8) & 67.4(2.3) & 3.01(10) & 76.7(6.1) & 1.22 \\
\hline  
Continuum  &  244(4) & 79(5) & 3.3(2) & 65(14) &\\ 
\hline
FLAG~\cite{FlavourLatticeAveragingGroupFLAG:2021npn} & 245--290 & 66--84 &  3.38(4) & 92.2(1.0) &\\
\end{tabular}  
\label{tab:summary_nf3v2}     
\end{table*}  

Besides the chiral condensates and
pion decay constants, we can also try to infer other sea information
such as the number of flavors and quark masses of the 
lattice gauge ensembles from the Dirac spectrum.

\begin{figure}[htbp]
    \centering
    \includegraphics[page=1, width=0.48\textwidth]{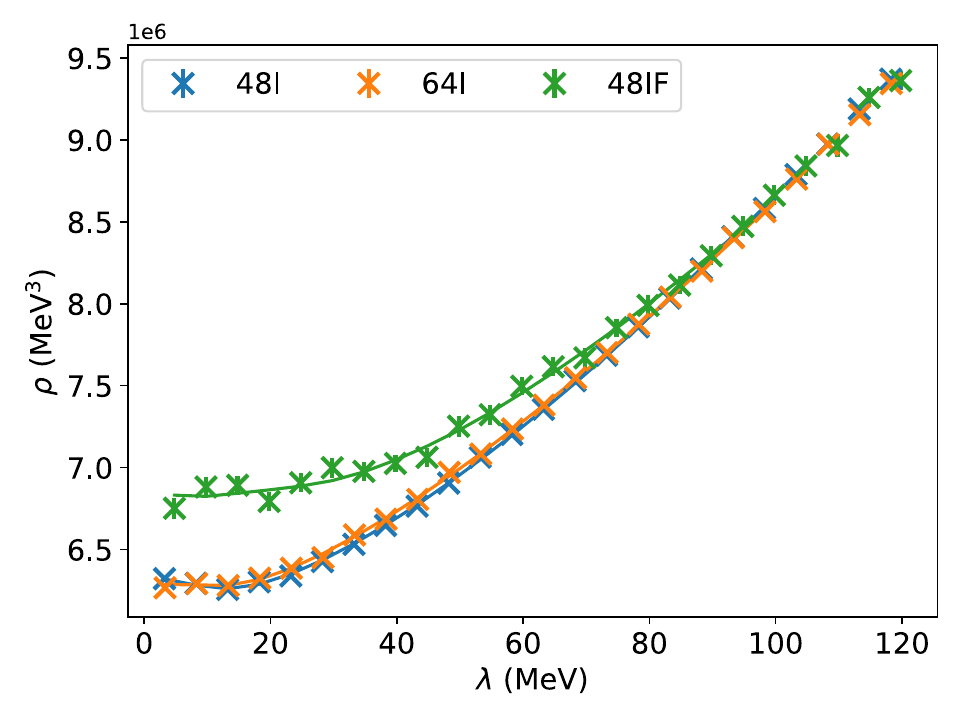}
    \caption{The $\rm SU(3)$ fitting in case 2 on the three lattices. The points are lattice data and the curves indicate the fitting.
    \label{fig:fits}}
\end{figure}

To this end, 
we redo the fits by using the $\rm SU(3)$ form and treating the strange quark mass $m_s$ and the light quark mass $m_l$ as free parameters in addition to $\Sigma_0$ and $F_0$. 
Since we now have more parameters, in principle we
can cover a bigger range of $\lambda$ in the fittings.
It turns out that, on the physical-point ensembles at $a=0.114$ and 0.084 fm, the smallest $\lambda$ we can reach is around 0.8 MeV
when setting the upper limit of the fit range to be 100 MeV. The data below $\lambda\sim0.8$ MeV possess significant finite-volume effects which can be seen in Fig.~\ref{fig:density}. 
The smallest $\lambda$ we can reach at $a=0.071$ fm is around 2.3 MeV which is easy to understand since its volume is small.
Fig.~\ref{fig:fits}
illustrates the fits and the results are collected in Tab.~\ref{tab:summary_nf3v2}. Since the finest lattice has unphysical
pion mass, we exclude the corresponding results in the table.
The $\chi^2/d.o.f.$'s are both close to one and 
it is actually remarkable that the $\rm SU(3)$ NLO $\chi$PT form can fit the lattice data from near-zero $\lambda$ to $\lambda$ greater than the strange quark mass.

Compared to the previous fittings, the new results of $\Sigma_0$ and $F_0$ are consistent
within 2 sigmas but they have much larger statistical uncertainties since more parameters are involved.
However, we do not quote the fittings in this section as a precise study.
The major point is to show the capability of revealing sea information
from the Dirac spectrum. 
Therefore, only statistical errors are included in the table.
We obtain $m_l=3.3(2)$ MeV which is consistent with the FLAG average 3.381(40) MeV
and $m_s=65(14)$ MeV which is less than 2-$\sigma$ away from the FLAG average 92.2(1.0) MeV~\cite{FlavourLatticeAveragingGroupFLAG:2021npn}. 
We emphasize here that the masses we get are
the masses used to generate the gauge ensembles.
Since the two ensembles we use are of physical pion mass,
so the values can be compared to the physical quark masses.
The results demonstrate that one can indeed obtain the
sea quark information from checking the lattice Dirac spectrum.
 
\section{Summary}

Based on the precise calculation of the spectral density $\rho(\lambda)$ of overlap Dirac operator on three lattice spacings with the statistical uncertainty at the 0.2\% level, we determine the chiral condensates in the $\rm SU(2)$ and $\rm SU(3)$ chiral limits to be $\Sigma=(265.4(0.5)(4.2)\ \textrm{MeV})^3$ and $\Sigma_0=(234.3(0.5)(25.8)\ \textrm{MeV})^3$ at $\overline{\textrm{MS}}$ scale 2 GeV.
We also determine the pion decay constants $F=84.1(1.9)(8.0)$ and $F_0=58.6(0.5)(10.0)$ MeV, respectively. The two uncertainties are the statistical one and the systematic one.
For $F_0$ and $\Sigma_0$,
the large systematic uncertainties are dominated by the effects of the change of the low-energy constant $L_6$.
In the discussion of non-perturbative renormalization,
we argue that the MOM scheme should be preferred as it is more reliable due to empirical ambiguities in fitting with the SMOM scheme.
  
The analysis of the Dirac spectrum also
allows us to determine the sea quark masses.
Physically, this is because the light quark mass makes $Z_v(\lambda,m_l)$ in the chiral form Eq.~(\ref{eq:def}) differ from unity due to the enhancement of the chiral log and changes the overall value of $\rho(\lambda)$, while the strange quark mass accounts for the $\lambda$ dependence of $\rho(\lambda)$.
Although with relatively large uncertainties,
the capability of obtaining sea information using the
Dirac spectrum is demonstrated.

\section*{Acknowledgment}
We thank the RBC and UKQCD collaborations for providing us their DWF gauge configurations. The calculations were performed using the GWU code~\cite{Alexandru:2011ee,Alexandru:2011sc} through the HIP programming model~\cite{Bi:2020wpt}. 
This work is partially supported by the Guangdong Major Project of Basic and Applied Basic Research No.\ 2020B0301030008. J.~L. is supported by the Natural Science Foundation of China under Grant No.\ 12175073 and No.\ 12222503, and the Natural Science Foundation of Basic and Applied Basic Research of Guangdong Province under Grant No.\ 2023A1515012712. A.~A. is supported in part by the U.S. Department of Energy, Office of Science, Office of Nuclear Physics under Grant No.\ DE-FG02-95ER40907. Y.~B. is supported in part by the National Natural Science Foundation of China (NNSFC) under Grant No. 12075253. T.~D. and K.~L. are supported by the U.S.\ DOE Grant No.\ DE-SC0013065 and DOE Grant No.\ DE-SC0023646 which is within the framework of the Quark-Gluon Tomography (QGT) Topical Collaboration. 
Y. Y is also supported by the NSFC grants No.\ 12293060, 12293062, and 12047503, the Strategic Priority Research Program of Chinese Academy of Sciences, Grant No.\ XDB34030303 and YSBR-101. The numerical calculation The numerical calculation were carried out on the ORISE Supercomputer, HPC Cluster of ITP-CAS,
and the Southern Nuclear Science Computing Center (SNSC).
This work also used Stampede time under the Extreme Science and Engineering Discovery Environment (XSEDE), which is supported by National Science Foundation Grant No.\ ACI1053575. We also used resources on Frontera at Texas Advanced Computing Center (TACC). We also thank the National Energy Research Scientific Computing Center (NERSC) for providing HPC resources that have contributed to the research results reported within this paper.

\bibliographystyle{apsrev4-1}
\bibliography{condensate.bib}

\vspace{10pt}
\appendix*

\end{document}